\documentclass[
 reprint,
amsmath,amssymb,
aps,
prl,
citeautoscript
]{revtex4-1}

\usepackage[Symbol]{upgreek}
\usepackage{amssymb}
\usepackage{mathrsfs}
\usepackage{graphicx}
\usepackage{dcolumn}
\usepackage{bm}
\usepackage{graphicx}
\usepackage{dcolumn}
\usepackage{bm}
\usepackage{xfrac}
\usepackage[english]{babel}
\usepackage{hyperref}
\usepackage{times}
\usepackage{amsfonts,amssymb,bm}
\usepackage{amsmath}
\usepackage{epsfig}
\usepackage{mathrsfs}
\usepackage{color,soul}
\usepackage{bbold}
\usepackage{subfig}
\usepackage{pstricks}
\usepackage{mathtools}
\usepackage{textcomp}
\usepackage{braket}
\usepackage{empheq}
\usepackage{amsmath}
\usepackage{amsthm}
\usepackage[babel]{csquotes}
\usepackage[font=small]{caption}
\usepackage[labelfont=bf,labelsep=quad,justification=raggedright]{caption}
\usepackage{epstopdf}

\begin{document}

\preprint{APS/123-QED}

\title{Measures of space-time non-separability of electromagnetic pulses}

\author{Yijie Shen$^{1,*}$, Apostolos Zdagkas$^{1}$, Nikitas Papasimakis$^{1}$, and Nikolay I. Zheludev$^{1,2,\dagger}$}
\affiliation{
	{$^{1}$Optoelectronics Research Centre \& Centre for Photonic Metamaterials, University of Southampton, Southampton SO17 1BJ, United Kingdom}\\
	{$^{2}$Centre for Disruptive Photonic Technologies, School of Physical and Mathematical Sciences and The Photonics Institute, Nanyang Technological University, Singapore 637378, Singapore}
}

\date{\today}
\begin{abstract}
Electromagnetic pulses are typically treated as space-time (or space-frequency) separable solutions of Maxwell's equations, where spatial and temporal (spectral) dependence can be treated separately. In contrast to this traditional viewpoint, recent advances in structured light and topological optics have highlighted the non-trivial wave-matter interactions of pulses with complex topology and space-time non-separable structure, as well as their potential for energy and information transfer. A characteristic example of such a pulse is the ``Flying Doughnut'' (FD), a space-time non-separable toroidal few-cycle pulse with links to toroidal and non-radiating (anapole) excitations in matter.  Here, we propose a quantum-mechanics-inspired methodology for the characterization of space-time non-separability in structured pulses. In analogy to the non-separability of entangled quantum systems, we introduce the concept of space-spectrum entangled states to describe the space-time non-separability of classical electromagnetic pulses and develop a method to reconstruct the corresponding density matrix by state tomography. We apply our method to the FD pulse and obtain the corresponding fidelity, concurrence, and entanglement of formation. We demonstrate that such properties dug out from quantum mechanics  quantitatively characterize the evolution of the general spatiotemporal structured pulse upon propagation. 
\end{abstract}

\maketitle

\textbf{Introduction} -- Electromagnetic pulses with tailored complex spatiotemporal structure are emerging as promising candidates for applications in communications~\cite{xie2018ultra}, particle acceleration~\cite{nie2018relativistic,hilz2018isolated}, laser machining~\cite{kerse2016ablation,penilla2019ultrafast,malinauskas2016ultrafast}, to name a few. The generation and diagnostics of such pulses is attracting growing interest from the  metamaterials~\cite{shaltout2019spatiotemporal,zhang2018space}, laser source~\cite{Can2020ultra,wright2017spatiotemporal,na2020ultrafast}, and nonlinear and topological photonics~\cite{shen2017gain,rego2019generation,dorney2019controlling,kondakci2017diffraction,shen2019optical} research communities. Typically, such pulses are treated as space-time (or equivalently space-frequency) separable solutions of Maxwell’s equations, that can be expressed as a product of a spatial mode and a temporal (or spectral) function, following the traditional separation of variables for solving partial differential equations~\cite{saleh2019fundamentals}. Since it is widely endorsed that any practical pulse is space-time separable, the space-time non-separability (STNS) is usually ignored for the sake of simplicity. However, the STNS can play a major role in the propagation dynamics~\cite{porras2002diffraction} and light matter interactions~\cite{hoff2017tracing}. A simple consequence of STNS existed in practical pulses is the separation of frequencies as the pulse propagates~\cite{porras2002diffraction,hoff2017tracing}, thus a pulse can be categorised as isodiverging or isodiffracting according to distribution of the spectral components that compose the pulse. The parameters of these monochromatic components can hence define the overall shape and characteristics of these pulses such as the change of the centre mass of the spectrum and the carrier envelope phase at focus which can then be tailored for the efficient control of attosecond processes~\cite{baltuvska2003attosecond}, chemical reactions~\cite{gordon2007effect} and ultrafast few-cycle pump-probe experiments~\cite{langer2016lightwave}. Furthermore, exact solutions of Maxwell’s equations for general waves of STNS is known to exist~\cite{donnelly1992method}, making their theoretical study more rigorous.

In 1983, Brittingham proposed the localized (e.g. non-diffracting) solutions to Maxwell’s equations termed focus wave modes~\cite{brittingham1983focus}, as the typical examples of STNS pulses. Although Brittingham’s modes required infinite energy, soon after, Ziolkowski showed that they arised STNS solutions to the scalar wave equation with moving complex sources~\cite{ziolkowski1985exact} and proposed that a superposition of such pulses leads to finite energy pulses termed ``electromagnetic directed-energy pulse trains''~\cite{ziolkowski1989localized}. Special cases of Ziolkowski’s solutions were studied by Hellwarth and Nouchi, who found closed-form expressions that describe single-cycle finite-energy STNS solutions to the homogeneous Maxwell’s equations. This family of pulses includes both linearly polarized pulses, termed ``pancakes''~\cite{feng1999spatiotemporal}, as well as pulses of toroidal symmetry, termed ``Flying  Doughnuts'' (FDs)~\cite{hellwarth1996focused}. The exotic FD pulses have holden promise of toroidal electrodynamics particularly in the contexts of nonradiating anapole configurations~\cite{baryshnikova2019optical,savinov2019optical}, topological information transfer~\cite{zdagkas2019singularities}, probing ultrafast light-matter interactions~\cite{raybould2016focused}, and toroidal excitations in matter~\cite{kaelberer2010toroidal,papasimakis2016electromagnetic}. Recently, it was demonstrated that the FD pulses can be generated by tailored metamaterials which can convert traditional few-cycle pulse into STNS pulses~\cite{papasimakis2018pulse,quevedo2019roadmap}.

Non-separability is also a quintessential property of quantum entanglement between particles: e.g. an entangled particle pair state cannot be expressed as the product of two single particle states, as a result the measurement of one particle effects the measurement outcome of another~\cite{horodecki2009quantum}. A typical example is the polarization-entangled photon pair where the polarization states of the two photons are non-separable. Over the past century, an extended toolbox has been developed that allows to quantify the non-separability of entangled states, including state tomography, density matrix, fidelity, linear entropy, concurrence, etc.~\cite{james2005measurement,toninelli2019concepts}. Recently, the tools of quantum mechanics were constructively applied not only to quantum physics but also to classical optics~\cite{konrad2019quantum,karimi2015classical,qian2015shifting,aiello2015quantum,forbes2019classically}. For example, the concept of quantum coherent state can be used to describe complicated laser modes~\cite{chen2003observation,chen2009transient,shen2018periodic,shen2018truncated}, that can mimic properties of high-dimensional quantum states~\cite{shen2020high}. The quantum Bell’s measure was also applied in classical optical coherence~\cite{kagalwala2013bell}. Many classical analogs of quantum states were realized in vortex beams such as Laughlin states~\cite{clark2020observation} and Shr\"odinger's cat states~\cite{liu2019classical}. Moreover, the vector vortex beams with spatially non-separable polarization can simulate the spin-orbital angular momentum entanglement~\cite{devlin2017arbitrary,mclaren2015measuring,giovannini2013characterization,forbes2019quantum}. These useful applications of quantum mechanics in classical optics have motivated the development of novel methods in optical (tele)communication~\cite{guzman2016demonstration,ndagano2017characterizing,ndagano2017creation}, cryptography~\cite{sit2017high}, optical computing~\cite{goyal2013implementing,goyal2015implementation,sephton2019versatile,d2020two}, metrology and sensing~\cite{d2013photonic,toppel2014classical,berg2015classically}.

\begin{figure*}[t!]
	\centering
	\includegraphics[width=\linewidth]{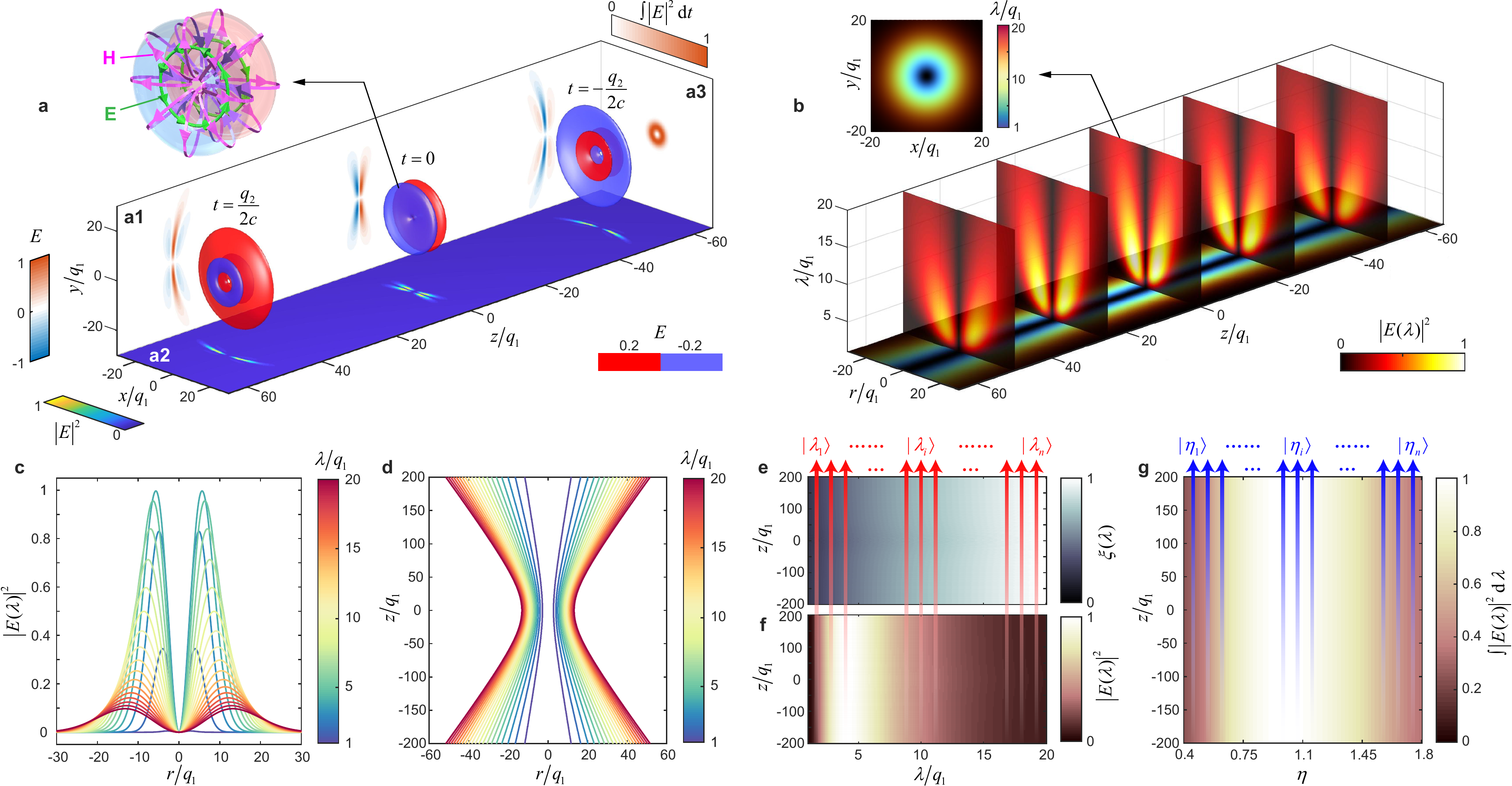}
	\caption{ \textbf{a}. \textbf{Spatiotemporal structure of the FD pulse.} The spatial isosurfaces of the electric field $\bm{E}(t,r,z)=\hat{\bm{\theta}}E(t,r,z)$ at different times of $t=0$ and $\pm q_2/(2c)$, at amplitude levels of $E=\pm0.2$; \textbf{a1}, the $y$-$z$ map of the instantaneous electric field $E$ of the FD pulse at $x=0$ for $t=0$ and $\pm q_2/(2c)$; \textbf{a2}, the $x$-$z$ map of the electric field intensity $|E|^2$ of the FD pulse at $x=0$, at $t=0$ and $\pm q_2/(2c)$; \textbf{a3}, The $x$-$y$ map of electric field intensity of the FD pulse integrated over all times at the $z=0$ plane $\int_{-\infty}^{\infty}|E(t,r,0)|^2\text{d}t$; the insert shows the electromagnetic vector structure of the FD pulse at focus ($z=0$). \textbf{b}. \textbf{Spectral structure of the FD pulse:} The colormaps in the $r$-$\lambda$ show $|\widetilde{E}(\lambda,r,z)|^2$ of the spectral components of the FD pulse at propagation distances $z$; The colormap in the $r$-$z$ plane shows a false color map of the positions; the inset shows an $x$-$y$ map of false color where the different positions of intensity maxima for different spectral components are revealed. \textbf{Isodiffraction of FD pulses}: \textbf{c}.  Profiles of radial distribution of normalized intensity $I(\lambda,r,z)=|\widetilde{E}(\lambda,r,z)|^2/\max_{(r,\lambda)}[|\widetilde{E}(\lambda,r,z)|^2]$ of different spectral components of the FD pulse at focus ($z=0$). Lines of different color represent monochromatic components of different wavelengths. \textbf{d}, The colour-coded traces of the positions where the intensity $I(\lambda,r,z)$ reach maxima for different wavelengths of the FD pulse; \textbf{e}. Ratio $\xi(\lambda,z)=r_\lambda/ r_{\lambda_n}$, where $r_\lambda$ is the radial position of the peak of the intensity $I(\lambda, r, z)$ of the monochromatic component at wavelength $\lambda$. Here,  the radius $r_{\lambda_n}$ is used for normalization and corresponds to the position of peak intensity for a given wavelength $\lambda_n$ of the FD pulse. \textbf{f}. Peak value of the intensity, $I_{\text{m}}(\lambda,z)=I(\lambda, r_{\lambda}, z)$, for each monochromatic component of the FD pulse as a function of propagation distance $z$. \textbf{g}, The normalized total field $I_0(r,z)=\int{I(\lambda, r, z)}\text{d}\lambda/\max_{r}[\int{I(\lambda, r, z)}\text{d}\lambda]$ plotted versus the value of $\eta(r,z)=r/r_{\text{max}}(z)$ at each propagation distance $z$, where $r_{\text{max}}(z)$ is the radius at which $I(r,z)$ reaches its maximum. Note the isodifraction property: $\xi(\lambda)$, $|\widetilde{E}_{\text{m}}(\lambda)|^2$, and $I(\eta)$ do not depend on $z$. The red and blue arrows demonstrate the positions of spectral and spatial states. Spectral states are represented by the trajectories of the electric field intensity maxima of the various monochromatic components, i.e. $r_{\lambda}(z)$, and spatial states are represented by the trajectories of prescribed positions $r(z)$ fulfilling the prescribed radial ratios of $\eta_i=r(z)/r_{\text{max}}(z)$.}
	\label{f1}
\end{figure*}

In this paper, we draw analogies between classical STNS waves and quantum entanglement and apply quantum methodologies to quantitatively characterize STNS of classical pulses. In particular, we present a state tomography approach to reconstruct the density matrix of space-time non-separable states. We apply our approach to general pulses with prescribed STNS, such as the FD pulse and a superposition of Laguerre-Gaussian (LG) modes. We demonstrate that the STNS measures introduced here allow to quantitatively characterize the evolution of the pulse spatio-spectral structure upon propagation. This work introduces a new toolkit for the characterization of a general family of pulses with various degrees of intrinsic space-time coupling, and provides a quantitative description of the spatiotemporal structure and the propagation dynamics of broadband pulses. The approach proposed here will lead to insights into light-matter interactions with ultrafast pulses and will find applications in spectroscopy, cryptography, and communications.

\textbf{Dynamics of FD pulse} -- FDs are few-cycle doughnut-like pulses with toroidal configuration of electric and magnetic fields. They exist both as transverse electric (TE) and transverse magnetic (TM) pulses. In the former case, the electric and magnetic fields are given by~\cite{hellwarth1996focused}:
\begin{equation}
\bm{E}=E_\theta\bm{\hat{\theta}}=-f_0i\sqrt{\frac{\mu_0}{\varepsilon_0}}\frac{r(q_1+q_2-2ict)}{\left[r^2+(q_1+i\tau)(q_2-i\sigma)\right]^3}\bm{\hat{\theta}}
\label{E}
\end{equation}
\begin{align}
\nonumber
\bm{H}=H_r\bm{\hat{r}}+H_z\bm{\hat{z}}=&f_0{i\frac{r(q_2-q_1-2iz)}{\left[r^2+(q_1+i\tau)(q_2-i\sigma)\right]^3}\bm{\hat{r}}}\\
&-f_0{\frac{r^2-(q_1+i\tau)(q_2-i\sigma)}{\left[r^2+(q_1+i\tau)(q_2-i\sigma)\right]^3}\bm{\hat{z}}}
\label{H}
\end{align}
where $\sigma=z+ct$, $\tau=z-ct$, $f_0$ is the amplitude parameter, $(q_1,q_2)$ represent the effective wavelength and Rayleigh range, respectively, and $(\bm{\hat{r}},\bm{\hat{\theta}},\bm{\hat{z}})$ are the three normalized basic vectors of cylindrical coordinates, respectively. In particular, the value of the ratio $q_2/q_1$ indicates whether the pulse is well-collimated $(q_2/q_1\gg1)$ or strongly focused. In the TE mode, the electric field is azimuthally polarized with no longitudinal or radial components, whereas the magnetic field is oriented along the radial and longitudinal directions with no azimuthal component (see Fig.~\ref{f1}a). Two different pulses can be constructed respectively from the real and imaginary parts of complex electromagnetic fields of Eqs.~(\ref{E}, \ref{H}), both types of which are exact solutions to Maxwell’s equations. The real part is single-cycle in the electric field and $1\frac{1}{2}$-cycle in the magnetic field at the focus ($z=0$), while the imaginary part is $1\frac{1}{2}$-cycle in the electric field and single-cycle in the magnetic field. Thus the real part is referred as the single-cycle pulse and the imaginary one as the $1\frac{1}{2}$-cycle pulse. Upon propagation, single-cycle ($1\frac{1}{2}$-cycle) transforms to the $1\frac{1}{2}$-cycle (single-cycle) pulse due to the Gouy phase shift~\cite{feng1998gouy}. The propagation dynamics of a single-cycle FD pulse ($q_2=100q_1$) is revealed by the isosurfaces of electric field at various times in Fig.~\ref{f1}a. Away from focus ($z=\pm q_2/2$), the pulse displays $1\frac{1}{2}$-cycle composed by a central bright doughnut and two darker toroidal lobes; at focus ($z=t=0$), the pulse is single-cycle with two equal-amplitude doughnuts corresponding to the two half-cycles of the pulse. 

\begin{figure}[t!]
	\centering
	\includegraphics[width=\linewidth]{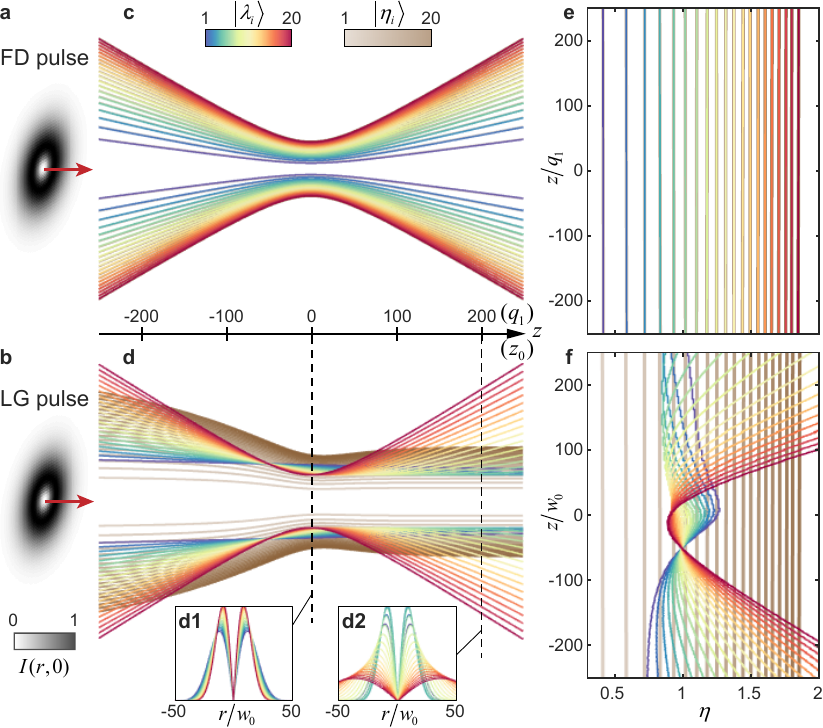}
	\caption{\textbf{a},\textbf{b}, The transverse  intensity patterns $I_0(r)|_{z=0}$ of the FD pulse (\textbf{a}) and a wide-band LG beam (see Supplementary Material Note~1) (\textbf{b}). \textbf{c},\textbf{d}, The propagation profiles of spectral ($|\lambda_i\rangle$) and spatial ($|r_i\rangle$) states of the FD beam (\textbf{c}) and wide-band LG beam (\textbf{d}). The two insets to panel~\textbf{d} show the spatial profiles of different wavelength components of the wide-band LG beam at two different propagation distances, $z=0$ (\textbf{d1}) and $z=200z_0$ (\textbf{d2}), respectively, where $z_0$ is $1/100$ of the Rayleigh length (averaged over all monochromatic components of the beam). \textbf{e},\textbf{f}, The $\eta$-$z$ map of spectral and spatial states of the FD beam (\textbf{e}) and wide-band LG beam (\textbf{f}). Note that the LG beam experiences dramatic distortion upon propagation, whereas the FD pulse profile remains invariant owing to its isodiffracting nature. In this illustration, we have considered 20 different spectral and spatial states, $|\lambda_i\rangle$ and $|\eta_i\rangle$ ($i=1,2,\cdots,20$), where wavelength values are set as $\lambda_i=iq_1$ and radial ratios are selected as $\eta_i=r_{\lambda_i}/r_{\text{max}}$.}
	\label{f2}
\end{figure}

\textbf{Space-spectrum ``entanglement''} -- Due to its spatiotemporal structure, the  FD pulse exhibits a frequency spectrum $\widetilde{E}(\lambda,r,z)$ with a complex spatial distribution covering a very broad spectral band \cite{zdagkas2019building}. Generally, all temporal properties can be fully characterized in the spectral domain, thus the STNS property can be equivalently interpreted by space-spectrum non-separability and the two terms will be used here interchangeably. The spatially dependent frequency spectrum of the FD pulse at various propagation distances is depicted in Fig.~\ref{f1}b. Here, the short-wavelength (bluish) components are always tightly confined close to the center of the doughnut, while the long-wavelength (reddish) components are located at the periphery of the pulse (see the insert in Fig.~\ref{f1}b). The spatial normalized intensity distribution, $I(\lambda,r,z)=|\widetilde{E}(\lambda,r,z)|^2/\max_{(r,\lambda)}[|\widetilde{E}(\lambda,r,z)|^2]$, of monochromatic components of different wavelengths $\lambda_i$ ($i=1,2,\cdots,n$) is depicted in Fig.~\ref{f1}c. The STNS in the FD pulse manifests as isodiffraction~\cite{feng2000spatiotemporal,zdagkas2019building}, ensuring that different spectral components of the pulse all diffract at the same rate, in other words, their spatial profiles experiences only transverse rescaling at a same rate upon propagation. To illustrate the isodiffracting nature of the FD pulse, we trace the radial position, $r_{\lambda_i}$, of the peak of the intensity of each wavelength upon propagation in Fig.~\ref{f1}d, that $I(\lambda_i,r_{\lambda_i},z)=\max_{r}[I(\lambda_i,r,z)]$. We introduce the dimensionless ratio $\xi=r_{\lambda_i}/r_{\lambda_n}$ of each trace, where the position of peak intensity of each monochromatic component is normalized to that of a given component at wavelength $\lambda_n$. In contrast to the radial positions of peak intensity (Fig.~\ref{f1}d), the ratio $\xi$ of each monochromatic component is propagation invariant (Fig.~\ref{f1}e). A similar propagation-invariant picture can be seen for the peak intensity value, $I_{\text{m}}(\lambda,z)=I(\lambda,r_{\lambda},z)$, of various wavelengths (Fig.~\ref{f1}f). To investigate the evolution of the transverse profile of total electric field intensity (integrated over the wavelength components), we introduce normalized radial positions $\eta=r(z)/r_{\text{max}}(z)$, where $r_{\text{max}}(z)$ is the position of the total electric field intensity in the transverse plane at propagation distance $z$. As shown in Fig.~\ref{f1}g, the normalized total intensity profile, $I_0(r,z)=\int I(\lambda, r, z)\text{d}\lambda/\max_{r}[\int I(\lambda, r, z)\text{d}\lambda]$, versus the normalized radius $\eta$ is also $z$-independent.  

The introduction of the radial position ratios, $\xi(\lambda)$, normalized radial coordinates, $\eta$, and the normalized electric field intensities, $I_{\text{m}}(\lambda)$ and $I_0(\eta,z)$, allow to highlight the propagation invariant characteristics of isodiffracting pulses, such as the FD. Indeed, in isodiffracting pulses, $\xi(\lambda)$, $I_{\text{m}}(\lambda)$, and $I_0(\eta)$ do not depend on the propagation distance, $z$. In contrast, a generic polychromatic beam (e.g. a wide-band superposed LG beam) is not expected to exhibit such propagation-invariant properties. Based on these properties, we can introduce two sets of states to describe STNS in broadband beams and pulses: 
(1) \textit{Spectral states} $|\lambda_i\rangle$ ($i=1,2,\cdots,n$) are (monochromatic) states of light of defined wavelength $\lambda_i$ and with defined radial position ($r_{\lambda_i}$) of peak intensity; (2) \textit{Spatial states} $|\eta_i\rangle$ are (generally polychromatic) states of light located at the position with defined radial ratio of $\eta_i=r/r_{\text{max}}$, where $r_{\text{max}}$ is the radial position at which the total intensity of the light field (e.g. the broadband beam or pulse) reaches its maximum. 
Generally the positions of the spectral and spatial states depend on propagation distance $z$. Here, for convenience we can also use the normalized radial position of $\eta=r/r_{\text{max}}$. Thus, at a transverse plane at propagation distance $z$, we can represent spectral state $|\lambda_i\rangle$ by the peak intensity $I_{\text{m}}(\lambda_i,z)$ at $\eta_{\lambda_i}(z)=r_{\lambda_i}(z)/r_{\text{max}}$. Similarly, a spatial state $|\eta_i\rangle$ can be represented by the intensity value $I_0(\eta_i r_{\text{max}},z)$ at normalized radial position $\eta_i$. The locations of spatial and spectral states, $\eta_i$ and $\eta_{\lambda_i}(z)$, respectively, define trajectories in the $\eta-z$ plane (see Fig.~\ref{f2}). For an arbitrary polychromatic beam, the trajectories of $\eta_i$ are always vertical lines in this plane, while $\eta_{\lambda_i}(z)$ can follow arbitrary trajectories. However, for an ideal isodiffracting pulse, both spectral and spatial states are represented by vertical trajectories reflecting the propagation invariance of the spatial and spectral intensity profile. Moreover, here, we choose sets of states in such way that spatial and spectral states are perfectly coincident, that is $\eta_{\lambda_i}(z)=\eta_i$ for isodiffracting pulses. 

The introduction of spatial and spectral sets of states allows to distinguish apparently similar broadband waves. As an example, we consider two doughnut-like pulses with different STNS, the FD pulse and a wide-band LG beam. Both pulses exhibit toroidal topology (see Figs.~\ref{f2}a and \ref{f2}b) and similar wide-band spectrum, but very different spatio-spectral structure and propagation dynamics as illustrated by the corresponding spatial and spectral states. For the FD pulse, the spectral states are coincident with the corresponding spatial states upon propagation, as Fig.~\ref{f2}c shows. In contrast, the wide-band LG beam is constructed by monochromatic LG modes, where each component is a space-time separable solution to the paraxial wave equation (see details of the wide-band LG beam construction in Supplementary Material Note~1). As a result, the corresponding spectral and spatial states are naturally separated (see Fig.~\ref{f2}d) and the spatio-spectral structure of the beam varies dramatically as it propagates. For example, at focus, long wavelength components are located close to the axis of the beam (Fig.~\ref{f2}d1), whereas away from focus they move to the periphery of the beam (Fig.~\ref{f2}d2). The difference between the isodiffracting FD and the broadband LG beam can be emphasized further in the $\eta$-$z$ plane. Here, as expected, the spectral states of the FD pulse (Fig.~\ref{f2}e) are $z$-invariant and coincident with the corresponding spatial states. On the other hand, the profile of the wide-band LG beam (Fig.~\ref{f2}f) suffers substantial distortion  as illustrated by the trajectories of the spectral states. This is a direct result of the non-coincidence of spectral and spatial states.

\begin{figure}[t!]
	\centering
	\includegraphics[width=0.96\linewidth]{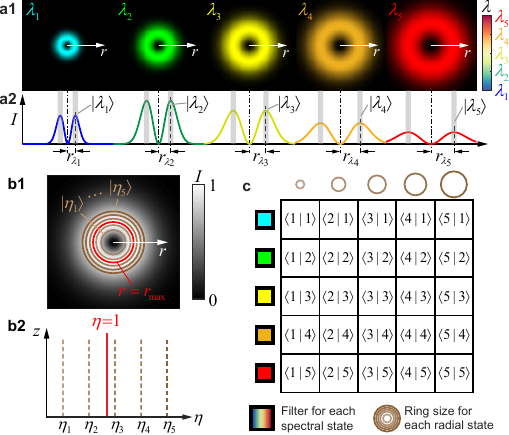}
	\caption{\textbf{a}, Procedure for experimental determination of spectral states $|\lambda_i\rangle$ ($i=1,2,\cdots,5$): \textbf{a1}, The transverse intensity profiles corresponding to different monochromatic components can be obtained by capturing an image of the pulse after propagation through spectral filters at selected wavelengths $\lambda_i$; \textbf{a2}, The radial distribution versus $r$ of the intensity patterns presented in (\textbf{a1}). The radii $r_{\lambda_i}$ mark the position at which the intensity of the monochromatic component of wavelength $\lambda_i$ reaches its maximum value. \textbf{b}, Procedure for experimental determination of spatial states $|\eta_i\rangle$ ($i=1,2,\cdots,5$): \textbf{b1}, The captured transverse profile of the total electric field intensity with the marked position of the recorded radius of $r=r_{\text{max}}$ ($\eta$=1) at which the total intensity reaches its maximum. Based on the recorded position of $\eta$=1, the positions of various spatial ($|\eta_i\rangle$) states can be recorded through radially scaling the position of $\eta$=1 by corresponding ratios of $\eta_i$; \textbf{b2}, Spatial states represented trajectories in the $\eta-z$ plane. As the trajectories of the spatial states are parallel to the $z$-axis in the $\eta$-$z$ map, each position of spatial state can be determined by the corresponding normalized radius $\eta_i=r/r_{\text{max}}$. \textbf{c}, Measurement matrix of state tomography of space-spectrum entangled states, where the inner products are noted as $\langle i|j\rangle=\langle\eta_i|\lambda_j\rangle$ ($i,j=1,2,\cdots,5$). The values of $\eta_i$ are selected so that the measurement matrix is diagonal for the ideal isodiffracting pulses. Here the sample number is set to 5 for spectral and spatial states for illustration purposes.}
	\label{fm}
\end{figure}
\begin{figure*}[t!]
	\centering
	\includegraphics[width=1\linewidth]{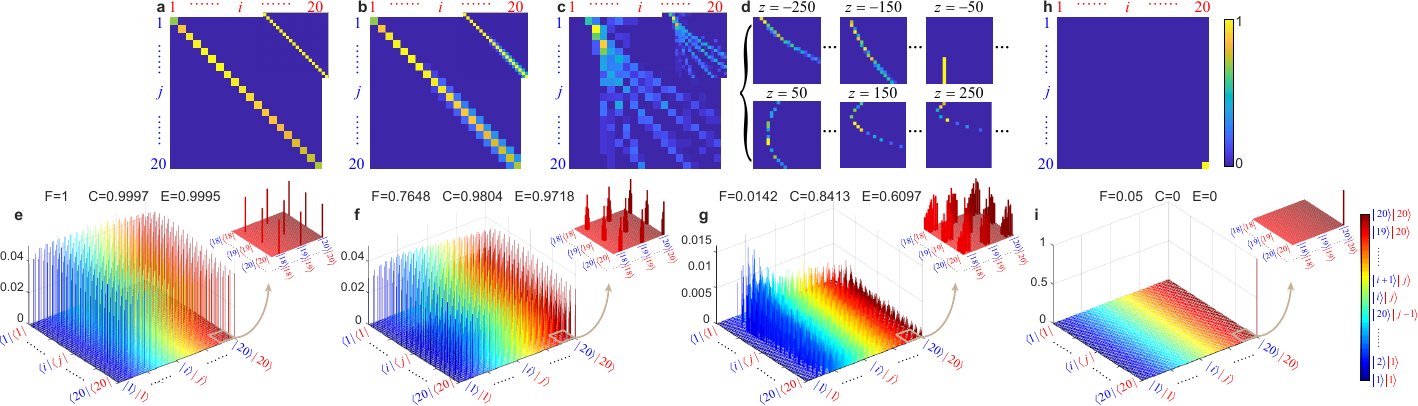}
	\caption{\textbf{a}-\textbf{c}, The results of quantum-analogous state tomography of the ideal FD pulse (\textbf{a}), the FD pulse with noise (\textbf{b}), and the wide-band LG pulse (\textbf{c}). The top-right insets to \textbf{a}-\textbf{c} are the corresponding results of the intensity-normalized measurements. In (c), the tomography matrix is obtained by averaging multiple measurements at various propagation distances from $z=-250$ to $z=250$ with step of $50$ (unit: $q_1$ for FD and $z_0$ for LG). \textbf{d}, The selected tomography matrices of the wide-band LG pulse at distances of $z=\pm250,\pm150,\pm50$ (unit: $z_0$). \textbf{e}-\textbf{g}, The reconstructed density matrices for the ideal FD pulse (\textbf{e}), the FD pulse with noise (\textbf{f}) and the wide-band LG pulse (\textbf{g}), with marked values of fidelity, concurrence, and EoF, respectively. \textbf{h},\textbf{i}, The tomography matrix (\textbf{h}) and density matrix (\textbf{i}) for a monochromatic beam. In all panels,  red and blue indices, $i$ and $j$, denote spatial  $|\eta_i\rangle$ and spectral states $|\lambda_j\rangle$. See the full dataset of tomography and density matrices of the wide-band LG beam at various propagation distances in Supplementary Material Note~3.}
	\label{f3}
\end{figure*}

To quantify STNS in doughnut-like pulses, we interpret the problem as a measurement of ``classical entanglement''~\cite{aiello2015quantum}, i.e. the non-separability of two classical fields. In our implementation, the classical fields of the spectral state $|\lambda_i\rangle$ and the spatial state $|\eta_i\rangle$ are represented as:
\begin{align}
{{\mathcal{E}}_{\lambda_i}}(r,z)&=\sqrt{I(\lambda_i,r,z)}H(r-\delta_{i-1}^{(\lambda)})H(\delta_{i}^{(\lambda)}-r)\label{lambda}\\
{{\mathcal{E}}_{\eta_i}}(r,z)&=\sqrt{I_0(r,z)}H(r-\delta_{i-1}^{(\eta)})H(\delta_{i}^{(\eta)}-r)\label{eta}
\end{align}
where $H(r)$ is the Heaviside step function $H(r)=1$ if $r>0$ and is zero otherwise, $\delta_{i}^{(\lambda)}=r_{\lambda_i}+{\Delta_i^{(\lambda)}}/{2}$ and $\delta_{i}^{(\eta)}=\eta_ir_{\text{max}}+{\Delta_i^{(\eta)}}/{2}$ for $i=1,2,\cdots,n-1$, $\delta_{0}^{(\lambda)}=\delta_{1}^{(\lambda)}-\Delta_1^{(\lambda)}$, $\delta_{0}^{(\eta)}=\delta_{1}^{(\eta)}-\Delta_1^{(\eta)}$, here $\Delta^{(\lambda)}_i$ ($\Delta^{(\eta)}_i$) for $i=1,2,\cdots,n-1$ is the distance between the positions of two adjacent spectral (spatial) states and $\Delta^{(\lambda)}_{i-1}=\Delta^{(\lambda)}_i$ ($\Delta^{(\eta)}_{i-1}=\Delta^{(\eta)}_i$) for $i=1$ and $n$, so that distributions of spectral (spatial) states are non-overlapping to each other. Both sets of spectral and spatial states fulfil the condition of orthogonal bases, $\langle\lambda_i|\lambda_j\rangle=\delta_{ij}$ and $\langle\eta_i|\eta_j\rangle=\delta_{ij}$, where $\delta_{ij}$ is the Kronecker delta. The inner product of two states is given by $\langle \eta_i|\lambda_j\rangle=\int \mathcal{E}_{\eta_i}\mathcal{E}_{\lambda_j}^*\text{d}r$. Here, the definitions of the classical fields ${{\mathcal{E}}_{\lambda_i}},{{\mathcal{E}}_{\eta_i}}$ have been introduced with respect to the radial coordinate, $r$, in order to clarify the experimental process for their retrieval. Equivalent definitions can be obtained in terms of the normalized radial coordinate, $\eta$, by substituting $r=\eta r_{\text{max}}$.

The classical fields of spectral and spatial states can be experimentally retrieved as follows. For a given spectral state $|\lambda_i\rangle$, the transverse profile of the monochromatic field propagating through a filter at the corresponding wavelength $\lambda_i$ can be recorded by a CCD camera at a given propagation distance $z$ (see Fig.~\ref{fm}a1). This allows to retrieve the peak position of the corresponding intensity $r_{\lambda_i}$ and calculate the field function by Eq.~(\ref{lambda}) (Fig.~\ref{fm}a2). For a spatial state $|\eta_i\rangle$, we should record the total intensity pattern (in the absence of spectral filters) at a propagation distance $z$, which allows to obtain the total intensity peak position $r_{\text{max}}$ (see Fig.~\ref{fm}b1). The corresponding field profile for the spatial state at $r=\eta_ir_{\text{max}}$ can then be calculated by Eq.~\ref{eta}. The values of $\eta_i$ are selected with reference to a perfectly isodiffracting pulse (such as the FD), so that the inner product $\langle\eta_i|\lambda_j\rangle$ is nonzero only if $i=j$ in this ideal STNS case.

Based on the above picture, the STNS is successfully translated into the non-separability of spectral and spatial states, which resembles the non-separability of entanglement, i.e. the classical entanglement. In quantum mechanics, there are plenty of mature techniques to quantitatively measure the non-separability for various kinds of high-dimensional entangled states, such as spin-to-orbital angular momentum entanglement~\cite{liu2020multidimensional}, energy-to-time entanglement~\cite{maclean2018direct}, and radial position-to-momentum entanglement~\cite{chen2019realization}. Based on the  analogous mathematical description and physical origin, we introduce the new concept of \textit{space-spectrum entangled state} that allows to quantitatively describe pulses with prescribed STNS, such as the FD, i.e. $|\psi\rangle=\sum_{i=1}^{n}c_i|\eta_i\rangle|\lambda_i\rangle$, where $c_{i}=\langle\eta_i|\lambda_i\rangle$. On the other hand, for a general pulse, the space-spectrum state is $|\psi\rangle=\sum_{i=1}^{n}\sum_{j=1}^{n}c_{i,j}|\eta_i\rangle|\lambda_j\rangle$, where $c_{i,j}=\langle\eta_i|\lambda_j\rangle$. Experimentally, spatiotemporal pulses can be precisely described by such states with a sufficient large number $n$ of measurements. We note that here we consider classical broadband beams and pulses as pure states (In quantum mechanics pure state means without a mixture of other states). Our approach can be readily expanded to mixed states, by examining e.g. pairs or triads of beams and pulses separated in space and/or time (akin to the implementation of mixed state in prior classical entanglement model~\cite{rafsanjani2015state}).

\textbf{Quantum-analogous measurement} -- In analogy with quantum state tomography, we can perform tomography measurements of the space-spectrum state of a spatiotemporal pulse, as Fig.~\ref{fm}c shows. Based on the definition of the spectral and spatial states adopted here, the tomography matrix for an isodiffracting pulse, such as the FD pulse should be diagonal, as shown in Fig.~\ref{f3}a, revealing space-spectrum entanglement. Importantly, the tomography matrix for isodiffracting pulses is diagonal at any transverse plane, i.e. it is propagation invariant. For comparison, we emulate a hypothetical experimentally generated FD pulse by adding noise (see Supplementary Material Note~2) into the ideal FD pulse, and calculate the corresponding tomography matrix as shown in Fig.~\ref{f3}b. Here, the presence of off-diagonal elements indicates that the spectral and radial states are slightly separated and that the pulse indeed deviates from the ideal one. On the other hand, for a wide-band LG beam without isodiffraction, the tomography results are propagation dependent. In this case, we average the tomography matrices (Fig.~\ref{f3}c) evaluated at various propagation distances (Fig.~\ref{f3}d). The tomography matrices evaluated at different transverse planes, as well as the averaged matrix, are non-diagonal indicating substantial deviation from isodiffracting propagation. Thus, the state tomography method introduced here allows to distinguish the type of STNS in broadband light fields. Indeed, both the FD pulse and the wide-band LG beam have degrees of STNS to some extend, however, only in the case of the FD does the ideal STNS leading to isodiffracting propagation. 

From the evaluated state tomography matrices, we can reconstruct the corresponding density matrices of the space-spectrum state, $\widetilde{\varrho}=|\widetilde{\psi}\rangle\langle\widetilde{\psi}|$ (where $|\widetilde{\psi}\rangle$ is the measured state). Results for the ideal FD, FD with noise, and wide-band LG pulses listed in Figs.~\ref{f3}e-\ref{f3}g, respectively. Importantly, knowledge of the density matrix allows to apply quantum tools to quantitatively characterize the properties of the pulse:

\textit{Fidelity.} In quantum mechanics, the fidelity is a measure of similarity of two quantum states, defined as $F=(\text{Tr}\sqrt{\sqrt{\rho_1}\rho_2\sqrt{\rho_1}})^2$, where $\rho_1$ and $\rho_2$ are the density matrices of the two states. If the target state is a pure state $|\psi_1\rangle$, the density matrix is given by $\rho_1=|\psi_1\rangle\langle\psi_1|$, and the fidelity is simplified to $F=\text{Tr}(\rho_1\rho_2)=\langle\psi_1|\rho_2|\psi_1\rangle$~\cite{james2005measurement}. Here, we set the target state as the ideal FD pulse $|\psi\rangle=\sum_{i=1}^{n}c_i|r_i\rangle|\lambda_i\rangle$. The fidelity of a measured state can then be calculated as $F=\langle\psi|\widetilde{\varrho}|\psi\rangle$, where $\widetilde{\varrho}$ is the density matrix of measured state. In our implementation, fidelity can quantitatively measure the degree of similarity to an ideal FD pulse taking values from 0 to 1. The result for the FD with noise is $F=0.7648$, which indicates high degree of similarity to the ideal FD, while in the case of the wide-band LG pulse fidelity approaches zero, $F=0.0142$.  Fidelity can be readily defined with respect to different reference pulses (e.g. linearly polarized STNS ``focused pancakes''~\cite{feng1999spatiotemporal}).

\textit{Concurrence.} In quantum mechanics, the concurrence is a continuous measure of non-separability of two-dimensional entangled states~\cite{james2005measurement}. This concept was also generalized for high-dimensional cases, usually called I-concurrence, defined by $C=\sqrt{2[1-\text{Tr}(\rho_A^2)]}$ where $\rho_A$ is the reduced density matrix~\cite{rungta2001universal}. For an arbitrary $d$-dimensional state, The concurrence is usually normalized as $C/\nu_d$ and takes values from 0 to 1 ($\nu_d=\sqrt{2(1-1/d)}$), indicating absence of entanglement (or pure separability) and strong non-separability (maximum entanglement), respectively. In our study, we use $d=20$ corresponding to the 20 spectral and spatial states. The results for the ideal FD, FD with noise, and wide-band LG beam are $C=0.9997$, $C=0.9804$ and $C=0.8413$, correspondingly. The FD pulse exhibits strong STNS with near-maximum ``entanglement'', while the wide-band LG pulse also exhibits substantial degree of STNS upon propagation owing to the mixing of the different monochromatic components.

\textit{Entanglement of formation.} In quantum mechanics, the entanglement of formation (EoF) is also a commonly encountered measure of quantum entanglement. EoF is calculated by the von~Neumann entropy of the reduced density matrix $E=-\text{Tr}[\rho_A\log_2(\rho_A)]$ and is typically normalized as $E/\log_2(d)$ in the $d$-dimensional case~\cite{wootters2001entanglement}. In contrast to concurrence, EoF is more sensitive to strong non-separability due to the convexity of entropic measures. The results for the ideal FD, FD with noise, and wide-band LG beam are $E=0.9995$, $E=0.9718$ and $E=0.6097$, respectively. The lower EoF of the  wide-band LG beam unveils that it exhibits weaker STNS. We note here that both EoF and concurrence quantify the degree of non-separability, yet the choice between the concurrence and EoF can be informed by the specific application at hand: EoF (concurrence) is better suited to distinguish between pulses with strong (weak) STNS (see Supplementary Material Note~4 for an example explanation).

\begin{table}
	\caption{\label{t1}%
		Parameter comparison of various kinds of pulse}
	\begin{ruledtabular}
		\begin{tabular}{ccccccc}
			\textrm{Pulse}&
			\textrm{Fid.}&
			\textrm{Conc.}&
			\textrm{EoF}&
			\textrm{N-Fid.\footnote{Here N-Fid. means the fidelity in intensity-normalized measurement. Similar meaning of N-Conc. and N-EoF. for concurrence and EoF.}}&
			\textrm{N-Conc.}&
			\textrm{N-EoF}\\
			\colrule
			Ideal FD & 1 & 0.9997 & 0.9995 & 1 & 1 & 1\\
			Noised FD & 0.7648 & 0.9804 & 0.9718 & 0.7533 & 0.9779 & 0.9683\\
			W. LG\footnote{The wide-band LG beam.} & 0.0142 & 0.8413 & 0.6097 & 0.0110 & 0.8541 & 0.6378\\
			M. LG\footnote{The monochromatic LG beam.} & 0.0410 & 0 & 0 & 0.0500 & 0 & 0\\
		\end{tabular}
	\end{ruledtabular}
\end{table}

We note that the ideal FD pulse exhibits very high values of concurrence and EoF (0.9997 and 0.9995), which indicates that the FD  is a near maximally entangled state. Here, the small deviation of the entanglement measures' values from unity is a result of different intensity levels at different spatial and spectral states. However, depending on the problem at hand, we can only care about the positions of states to measure their separability independent of intensity, then we could regard the FD pulse as a perfect maximally entangled state. In such a case, we can use an intensity-normalized calculation of the inner product 
$\langle \eta_i|\lambda_j\rangle=\int \mathcal{E}_{\eta_i}\mathcal{E}_{\lambda_j}^*\text{d}r/\left(\int  |\mathcal{E}_{\eta_i}|\text{d}r\int|\mathcal{E}_{\lambda_j}|\text{d}r\right)$ during state tomography. For the ideal FD pulse, the intensity-normalized measurement results into an identity tomography matrix, while fidelity, concurrence, and EoF are all unity. The results of this modified measurement for the ideal FD, the FD with noise, and the wide-band LG beam are inserted in Figs.~\ref{f3}a-\ref{f3}c, correspondingly. 

As an extreme case of space-time separable wave, we consider a monochromatic LG beam. The corresponding results of tomography and density matrix are presented in Figs.~\ref{f3}h and \ref{f3}i, exhibiting only a single non-zero element. As a separable state, it can be expressed in the form  $|\eta_n\rangle|\lambda_n\rangle$, resulting in null values of concurrence and EoF. We summarize the results of fidelity and entanglement measures in Table~\ref{t1}, for both normalized and non-normalized intensity measurements. 

\textbf{Discussion} -- We have established a toolkit of quantum-analogous methods to effectively characterize STNS in general electromagnetic beams or pulses, which can not only evaluate the type, but also quantify the strength of the non-separability. The approach is straightforward and can be easily applied to experimental measurements. Measures such as fidelity, concurrence, and EoF borrowed from quantum mechanics can fully quantify the STNS of an general pulse.
  
While here we focus on fidelity, concurrence and EoF, a much wider set of quantities has been developed to measure the purity and quality of quantum states, e.g. linear entropy, negativity, Bell parameter, Greenberger-Horne-Zeilinger parameter, to name a few. Hence, this set can be mined to further characterize broadband electromagnetic waves with space-time or even more exotic forms of non-separability. For example, the linear entropy $S=1-\text{Tr}(\rho^2)$, where $\rho=\sum_jp_j|\psi_j\rangle\langle\psi_j|$ is the density matrix of a measured state and $p_j$ the coefficient of $j$-th mixed state, quantifies how close a quantum state is to a pure ($S=0$) or a maximally mixed ($S\to1$) state~\cite{peters2004mixed}. In this paper, we only consider waves that are represented by pure states, thus the linear entropy of such waves should always be 0. However, the linear entropy would be very useful, if we consider systems comprising multiple beams. Such a case would be of great interest as we can use the entropy of sets of multiple pulsed beams to encode information, enabling novel applications in high-capacity and encrypted communications by the STNS of pulses.

Space-time non-separable pulses provide unusual and largely unexplored degrees of freedom in structuring light that are yet to be exploited. As such, there is growing interest in the generation and control of high-quality space-time non-separable pulses. Our method provides the practical quantitative tools for the generation design, optimization, characterization and detection of such complex pulses, as well as for the study of their light-matter interactions. These key capabilities for taming and exploiting spatiotemporally structured pulses will lead to novel applications in ultra-high-capacity communications, high-security encryption, topology- and quantum-analogous systems, and metrology that require the manipulation of an increasing number of degrees of freedom. 

\textbf{Acknowledgements} -- The authors acknowledge the supports of the MOE Singapore (MOE2016-T3-1-006), the UKs Engineering and Physical Sciences Research Council (grant EP/M009122/1, Funder Id: http://dx.doi.org/10.13039/501100000266), the European Research Council (Advanced grant FLEET-786851, Funder Id: http://dx.doi.org/10.13039/501100000781), and the Defense Advanced Research Projects Agency (DARPA) under the Nascent Light Matter Interactions program. Yijie Shen thanks Andrew Forbes, Bienvenu Ndagano and Isaac Nape in University of the Witwatersrand for useful discussions.

{$^*$y.shen@soton.ac.uk}

{$^\dagger$zheludev@soton.ac.uk}


\bibliography{apssamp}
%

\end{document}